\documentclass[conference]{IEEEtran}
\usepackage{stfloats}
\usepackage{amssymb}
\usepackage{graphicx}
\usepackage{amsmath}
\begin{document}
%
\title{A Low Complexity Algorithm and Architecture for Systematic Encoding of Hermitian Codes}
\author{\authorblockN{Rachit Agarwal\authorrefmark{1}\authorrefmark{2},
Ralf Koetter\authorrefmark{3} and 
Emanuel M. Popovici\authorrefmark{1}\authorrefmark{4}}
\authorblockA{\authorrefmark{1} Department of Microelectronic Engineering, University College Cork, Cork, Ireland\\}
\authorrefmark{2} Microelectronics Application Integration Group, Tyndall National Institute, Cork, Ireland\\
\authorrefmark{3} Institute for Communications Engineering, Technische Universitaet, Muenchen, Germany\\
\authorrefmark{4} Claude Shannon Institute for Discrete Mathematics, Coding and Cryptography, Ireland\\
Email: rachit.agarwal@ue.ucc.ie, ralf.koetter@tum.de, e.popovici@ucc.ie}
\maketitle
\begin{abstract}
We present an algorithm for systematic encoding of Hermitian codes. For a Hermitian code defined over $GF(q^2)$, the proposed algorithm achieves a run time complexity of $O(q^2)$ and is suitable for VLSI implementation. The encoder architecture uses as main blocks $q$ varying-rate Reed-Solomon encoders and achieves a space complexity of $O(q^2)$ in terms of finite field multipliers and memory elements.
\end{abstract}
\section{Introduction}
Algebraic-Geometric (AG) codes \cite{goppa} offer desirable properties such as large code lengths over small finite fields, the potential to find a large selection of codes and good error-correction at high code rates \cite{perform}. In recent years, an important class of one-point AG codes, called Hermitian codes, has been frequently discussed \cite{generalrs}-\cite{structure}.

For a Hermitian code defined over $GF(q^2)$, a brute-force way to design an encoder is to multiply the information vector by a generator matrix. The space complexity of a serial-in serial-out architecture for this systematic encoder is $O(q^5)$ in terms of finite field multipliers and $O(q^3)$ in terms of memory elements. The encoder requires $2n$ clock cycles to generate a codeword of length $n$, thus, the latency is $n$.

By considering a Hermitian code as a superposition of several generalized Reed Solomon (RS) codes, an encoding scheme is introduced in \cite{generalrs2}. In \cite{shen}, an encoding algorithm by forming a bivariate information polynomial and evaluating this polynomial at every finite rational point on the Hermitian curve is proposed. However, both such schemes are nonsystematic and involve the evaluation of bivariate polynomials at $n$ finite rational points, thus, they may not have efficient hardware architecture for implementations. 

A computationally efficient approach for systematic encoding was proposed in \cite{grobner}. A serial-in serial-out architecture for this approach was proposed in \cite{serialinout}. This architecture requires $n$ clock cycles to encode a codeword of length $n$. The space complexity for this architecture is $O(q^3)$, both in terms of finite field multipliers and memory elements.

In this paper, we present an algorithm for systematic encoding and syndrome computation of Hermitian codes. We give an outline for the encoder architecture, which uses $q$ varying-rate RS encoders as main blocks and requires $n^{2/3}$ clock cycles for encoding a codeword of length $n$. The space complexity of the architecture is $O(q^2)$ in terms of both, memory elements and finite field multipliers. 
%
%
%
%
\section{Hermitian Codes and Syndrome Computation}
\label{hcasc}
We consider codes from a Hermitian curve
\begin{displaymath}
\chi : x^{q+1} = y^q + y
\end{displaymath}
over a finite field $\mathbb{F}_{q^2}$. The space $L(mP_\infty)$ consists of all functions on $\chi$ that have a pole of multiplicity at most $m$ only at the unique point at infinity. For $L(mP_\infty)$, we choose a basis
\begin{displaymath}
L(mP_\infty) = \langle{x^ay^b : aq + b(q+1) \leq m,\ 0 \leq a,\ 0 \leq b < q}\rangle
\end{displaymath}

Let $y_0$ be an element of $\mathbb{F}_{q^2}$ such that $y_0 + {y_0}^q = 1$. The affine rational points on $\chi$ are of the form
\begin{displaymath}
P_{\alpha, \beta} = (\alpha, \alpha^{q+1}(y_0 + \beta) + \delta(\alpha)\beta),
\end{displaymath}
where $\delta$ is the Kronecker-delta and $\alpha$ and $\beta$ represent arbitrary elements in $\mathbb{F}_{q^2}$ and $\mathbb{F}_q$ respectively.

Let $\epsilon$ be a primitive element in $\mathbb{F}_{q^2}$ and let $\gamma$ be a primitive element in $\mathbb{F}_q$. We label the positions in a codeword by the corresponding elements $\alpha$ = $\epsilon^i$, $\beta$ = $\gamma^j$ and we thus naturally consider a codeword as a $q \times q^2$ matrix ${\bf{c}}$. Occasionally we will index elements in this array by elements of the fields $\mathbb{F}_{q^2}$ and $\mathbb{F}_{q}$, otherwise we index starting with $0$.

A Hermitian code $C(m)$ is defined as 
\begin{displaymath}
\{{\bf{c}} \in \mathbb{F}_{q^2}^{q^3}: \sum_{\alpha \in \mathbb{F}_{q^2}} \sum_{\beta \in \mathbb{F_q}} {\bf{c}}_{\beta,\alpha} f(P_{\alpha,\beta}) = 0,\ \forall f \in L(mP_\infty)\}
\end{displaymath}

For an in-depth treatment of AG codes we refer to \cite{book1}. Throughout this paper we consider $m$ and thus the Hermitian code as being fixed.

Given a $q \times q^2$ matrix ${\bf{r}}$ we can check if ${\bf{r}}$ is a codematrix in a Hermitian code by computing the syndromes 
\begin{displaymath}
S_{a,b}({\bf{r}}) = \sum_{\alpha \in \mathbb{F}_{q^2}} \sum_{\beta \in \mathbb{F}_{q}} {\bf{r}}_{\beta,\alpha}(x(P_{\alpha,\beta}))^a (y(P_{\alpha,\beta}))^b
\end{displaymath}
${\bf{r}}$ is a code-matrix iff $S_{a,b}({\bf{r}})$ is zero for all $x^a y^b \in L(mP_\infty)$. Substituting the explicit form of the points we get
\begin{displaymath}
S_{a,b}({\bf{r}}) = \sum_{\alpha \in \mathbb{F}_{q^2}} \sum_{\beta \in \mathbb{F}_{q}} \alpha^a (\alpha^{(q+1)}(y_0 + \beta) + \delta(\alpha)\beta)^b {\bf{r}}_{\beta,\alpha}
\end{displaymath}
These equations can further be developed to give specific forms as shown in (\ref{eqn_dbl_x}) and (\ref{eqn_dbl_y}).
\begin{figure*}[b]
\normalsize
\hrulefill
\begin{equation}
\label{eqn_dbl_x}
S_{a,b}({\bf{r}}) = \left\{ \begin{array}{ll}
\sum_{\alpha \in \mathbb{F}_{q^2}\backslash\{0\}} \sum_{\beta \in \mathbb{F}_{q}} \alpha^a \alpha^{b(q+1)}{(y_0 + \beta)}^b {\bf{r}}_{\beta,\alpha} & \textrm{a $\neq$ 0}\\
\sum_{\alpha \in \mathbb{F}_{q^2}} \sum_{\beta \in \mathbb{F}_{q}} (\alpha^{b(q+1)}(y_0 + \beta)^b + \delta(\alpha)\beta^b) {\bf{r}}_{\beta,\alpha} & \textrm{a = 0}
\end{array} \right.
\end{equation}
\begin{equation}
\label{eqn_dbl_y}
S_{a,b}({\bf{r}}) = \left\{ \begin{array}{ll}
\sum_{i=0}^{q^2-2} \epsilon^{i(a+b(q+1))}(y_0^b{\bf{r}}_{0,i} + \sum_{j=0}^{q-2} {(y_0 + \gamma^j)}^b {\bf{r}}_{j+1,i}) & \textrm{a $\neq$ 0}\\
\sum_{i=0}^{q^2-2} \epsilon^{ib(q+1)}(y_0^b{\bf{r}}_{0,i} + \sum_{j=0}^{q-2} {(y_0 + \gamma^j)}^b {\bf{r}}_{j+1,i}) + \sum_{j=0}^{q-2} \gamma^{jb} {\bf{r}}_{j+1,q^2-1} & \textrm{a = 0}
\end{array} \right.
\end{equation}
\vspace*{4pt}
\end{figure*}
From the structure of (\ref{eqn_dbl_y}), it comes naturally to define a matrix as in (\ref{defA}) to convert the expression into a matrix multiplication.
\begin{figure*}
\begin{equation}
\label{defA}
\mathbf{A} =
\left( \begin{array}{ccccc}
(y_0 + 0{)^{0}} & (y_0 + \gamma^0)^{0} & (y_0 + \gamma^1)^{0} & \ldots & (y_0 + \gamma^{q-2})^0\\
(y_0 + 0{)^{1}} & (y_0 + \gamma^0)^{1} & (y_0 + \gamma^1)^{1} & \ldots & (y_0 + \gamma^{q-2})^1\\
(y_0 + 0{)^{2}} & (y_0 + \gamma^0)^{2} & (y_0 + \gamma^1)^{2} & \ldots & (y_0 + \gamma^{q-2})^2\\
\vdots & \vdots & \vdots & \ddots & \vdots\\
(y_0 + 0{)^{q-1}} & (y_0 + {\gamma^0})^{q-1} & (y_0 + \gamma^1)^{q-1} & \ldots & (y_0 + {\gamma^{q-2}})^{q-1}\\
\end{array} \right)
\end{equation}
\vspace*{4pt}
\end{figure*}
Similarly we define a matrix $A'$ as
\begin{displaymath}
\mathbf{A'} =
\left( \begin{array}{ccccc}
1 & (\gamma^0)^{0} & (\gamma^1)^{0} & \ldots & (\gamma^{q-2})^0\\
0 & (\gamma^0)^{1} & (\gamma^1)^{1} & \ldots & (\gamma^{q-2})^1\\
0 & (\gamma^0)^{2} & (\gamma^1)^{2} & \ldots & (\gamma^{q-2})^2\\
\vdots & \vdots & \vdots & \ddots & \vdots\\
0 & (\gamma^0)^{q-1} & (\gamma^1)^{q-1} & \ldots & (\gamma^{q-2})^{q-1}\\
\end{array} \right)
\end{displaymath}

For later use we give here the following Lemma.
\subsection{Lemma 1}
{\it{The $l \times l$ submatrices of $A$ consisting of the elements indexed by $i,j$ and that of $A'$ consisting of the elements indexed by $j,i$, $ i = q-l,\ q-l+1, \dots, q-1$, $j = 0,\ 1, \dots, l-1$ are non-singular.}}\\
\begin{proof}
This lemma follows in both cases from the properties of Vandermonde matrices.
\end{proof}
\par
It will be convenient to define an array $\overline{A}$ of matrices of type $A$ and $A'$.
\begin{displaymath}
\overline{A} = (A_0, A_1, \dots, A_{q^2-1}),\
A_i = \left\{ \begin{array}{ll}
A' & \textrm{i = $q^2$-1 }\\
A & \textrm{otherwise}
\end{array} \right.
\end{displaymath}

Given a $q \times q^2$ array ${\bf{r}}$ with columns ${\bf{r}}_j$, we define a $q \times q^2$ matrix ${\bf{\tilde{r}}}$ with columns ${\bf{\tilde{r}}}_j$ as
\begin{displaymath}
\tilde{{\bf{r}}}_j = A_j{\bf{r}}_j
\end{displaymath}

One of the main ingredients in both the syndrome calculation and a systematic encoding is the use of techniques for cyclic codes which are extended by one extra position. Let $\hat{a}(b) = \lfloor (m-(q-1-b)(q+1))/q \rfloor = max(a : x^a y^b \in L(mP_\infty))$.

\subsection{Definition 1}
{\it{Let an ordered set $\Re = \{\xi_0, \xi_1, \dots\}$ of elements from $\mathbb{F}_{q}$ be given. We define the code $EC(\Re,q)$ as
\begin{displaymath}
\{c \in \mathbb{F}_q^q : \sum_{i=0}^{q-2}c_i\xi^i = 0,\ \forall \xi \in \Re \backslash \{\xi_0\},\ c_{q-1} + \sum_{i=0}^{q-2}c_i{\xi_0}^i = 0\}
\end{displaymath}}}

For the natural indexing of elements in $\mathbb{F}_{q}$ and $\mathbb{F}_{q^2}$ induced by $\gamma$ and $\epsilon$ we have the following Lemma.

\subsection{Lemma 2}
{\it{
Let a $q \times q^2$ matrix ${\bf{r}}$ be given. The matrix ${\bf{r}}$ is a code matrix in the Hermitian code $C(m)$ iff the $i$th row of $\tilde{{\bf{r}}}$ is a codeword in $EC((\epsilon^{0+i(q+1)}, \epsilon^{1+i(q+1)}, \dots, \epsilon^{\hat{a}(i)+i(q+1)}), q^2)$.}}

\begin{proof}
The proof follows immediately from the syndrome definition.
\end{proof}

Codes of type $EC((\epsilon^{i(q+1)}, \epsilon^{1+i(q+1)}, \dots, \epsilon^{\hat{a}(i)+i(q+1)}), q^2)$ will play a central roll in the sequel. We define codes $E_i$ as 
\begin{displaymath}
E_i = EC((\epsilon^{0+i(q+1)}, \epsilon^{1+i(q+1)}, \dots, \epsilon^{\hat{a}(i)+i(q+1)}), q^2)
\end{displaymath}

From Lemma 2 we can derive an efficient way to compute the syndrome for a Hermitian code. Given a received matrix {\bf{r}} we obtain a matrix $\tilde{{\bf{r}}}$ with columns $\tilde{{\bf{r}}}_j$ = $A_j{\bf{r}}_j$. 
%

Given $\tilde{{\bf{r}}}$ we can easily solve the task of computing syndromes provided we can compute the corresponding syndromes for codes $E_i,\ i = 0, 1, \dots, q-1$. 
%
%
\section{Systematic Encoding}
\label{sysen}
The idea behind the systematic encoding of Hermitian codes is to use the well known techniques for the systematic encoding of cyclic codes. Lemma 2 almost immediately gives a nonsystematic encoding procedure for Hermitian codes. To this end let $\tilde{{\bf{r}}}$ be a $q \times q^2$ matrix such that the $j$th row of $\tilde{{\bf{r}}}$ is a codeword in $E_j$. It follows from Lemma 2 that we can obtain a code-matrix for a Hermitian code by multiplying the columns of $\tilde{{\bf{r}}}$ with matrices $A^{-1}$ and $A'^{-1}$ respectively. We can obtain such a matrix $\tilde{{\bf{r}}}$ using eg. systematic encoding procedures for codes of type $E_j,\ j = 0,\dots,q-1.$

We will need $A^{-1}$ and $A'^{-1}$.
\subsection{Lemma 3} 
{\it{The matrices $A$ and $A'$ have inverses given in (\ref{invA}) and (\ref{invAi}).
\begin{figure*}
\begin{equation}
\label{invA}
\mathbf{A^{-1}} =
\left( \begin{array}{cccc}
1- (y_0 + 0{)^{q-1}} & (y_0 + 0{)^{q-2}} & \ldots & (y_0 + 0{)^0}\\
1- (y_0 + 1{)^{q-1}} & (y_0 + 1{)^{q-2}} & \ldots & (y_0 + 1{)^0}\\
1- (y_0 + \gamma{)^{q-1}} & (y_0 + \gamma{)^{q-2}} & \ldots & (y_0 + \gamma{)^0}\\
\vdots & \vdots & \ddots & \vdots\\
1- (y_0 + {\gamma^{q-2}}{)^{q-1}} & (y_0 + {\gamma^{q-2}}{)^{q-2}} & \ldots & (y_0 + {\gamma^{q-2}}{)^0}\\
\end{array} \right)
\end{equation}
\vspace*{4pt}
\end{figure*}

\begin{figure*}
\begin{equation}
\label{invAi}
\mathbf{A'^{-1}} =
\left( \begin{array}{cccccc}
1 & 0 & 0 & \ldots & 0 & -1\\
0 & -(1{)^{q-2}} & -(1{)^{q-3}} & \ldots & -(1{)^1} & -1\\
0 & -(\gamma{)^{q-2}} & -(\gamma{)^{q-3}} & \ldots & -(\gamma{)^1} & -1\\
\vdots & \vdots & \vdots & \ddots & \vdots\\
0 & -({\gamma^{q-2}}{)^{q-2}} & -({\gamma^{q-2}}{)^{q-3}} & \ldots & -({\gamma^{q-2}}{)^1} & -1\\
\end{array} \right)
\end{equation}
\vspace*{4pt}
\hrulefill
\end{figure*}}}

\begin{proof}
The inverse of $A'$ is straight forward to verify. We only show the inverse of $A$. The rows of $A^{-1}$ and the columns of $A$ may be thought of as being indexed by elements of $\mathbb{F}_q$. Let $C$ be the matrix obtained as $C = A^{-1}A$. The entry $C_{i,j}$ is thought of as being indexed by $\mu,\nu \in \mathbb{F}_q$.
\begin{displaymath}
\begin{array}{ll}
C_{\mu, \nu} = & (1-(y_0 + \mu)^{q-1}(y_0+\nu)^0 - (y_0 + \mu)^{q-2}(y_0 + \nu)^1\\ 
&\dots -(y_0+\mu)^0(y_0+\nu)^{q-1})\\
&\\
& = 1 - \sum_{i=0}^{q-1}(y_0+\mu)^{q-1-i}(y_0+\nu)^{i}\\
&\\
& = 1 - \dfrac{(y_0 + \mu)^q-(y_0+\nu)^q}{y_0 + \mu - (y_0 + \nu)}\\
\end{array}
\end{displaymath}
\begin{displaymath}
1 - (\mu - \nu)^{q-1} = \left\{ \begin{array}{ll}
1 & \textrm{$\mu = \nu$}\\
0 & \textrm{$\mu \neq \nu$}
\end{array} \right.
\end{displaymath}
\end{proof}
We note that we are entirely free to choose "virtual information symbols" in matrix $\tilde{{\bf{r}}}$. Let a sequence of information symbols be given that are to be encoded systematically in a codeword of a Hermitian code. The trick in obtaining a systematic encoding procedure is to choose the information symbols in $\tilde{{\bf{r}}}$ so that the mapping with $A^{-1}$ and $A'^{-1}$ respectively, gives the primary information symbols that we really want to encode.

Before we derive a systematic encoding procedure for Hermitian codes, we treat a somewhat simpler case, which will elucidate the idea of systematic encoding. Let $\hat{C}$ be a code on a Hermitian curve defined as
\begin{displaymath}
\begin{array}{ll}
\hat{C} = & \{c \in \mathbb{F}_{q^2}^{q^3}\ :\ S_{a,b}(c) = 0,\\
&\ a = 0, 1, \dots, \hat{a} < q^2-1,\,\ b = 0, 1, \dots, q-1\} \\
\end{array}
\end{displaymath}
The code $\hat{C}$ has dimension $(q^2-\hat{a}-1)q$. The following algorithm may be used for systematic encoding of code $\hat{C}$.
\subsection{Algorithm 1}
{\it{
\begin{enumerate}
\item Write the $(q^2-\hat{a}-1)q$ information symbols in an array $d$ of size $q \times (q^2 - \hat{a} - 1)$
\item Compute $\hat{\bf{r}} = Ad^T$
\item Encode the $i$th row of $\hat{\bf{r}}$ independently in a systematic way into codewords of the code 
\begin{displaymath}
EC((\epsilon^{0+i(q+1)}, \epsilon^{1+i(q+1)}, \dots, \epsilon^{\hat{a}(i)+i(q+1)}), q^2)
\end{displaymath}
Denote the resulting $q \times q^2$ matrix with $\hat{\bf{r}}'$.
\item Compute columns $c_i$ = $A_i^{-1}\hat{\bf{r}}'_i$
\end{enumerate}
}}
Algorithm 1 yields a systematic encoding procedure for the code $\hat{C}$ because $c$ is a code matrix by Lemma 2 and the first $(q^2 - \hat{a} - 1) \times q$ symbols are the original information symbols. The first $(q^2 - \hat{a} - 1)$ columns of $d$ determine the first $(q^2 - \hat{a} - 1)$ columns of $\hat{\bf{r}}$. It is the first $(q^2 - \hat{a} - 1)$ columns of $\hat{\bf{r}}$ that contain the virtual information symbols for the encoding of the cyclic codes.

The situation for Hermitian codes is complicated by the fact that the codes $E_i$ have different rates. Thus at some instance of the algorithm we have to process the columns that are in one part determined by information symbols and the other part is determined by redundancy symbols generated by the systematic encoders of the codes $E_i$. For simplicity, we will restrict our attention to codes $C(m)$ of dimension $k$ that is less than $(q^3 - g - q)$.

Let $\phi_i\ :\ \mathbb{F}_{q^2}^{q^2-\hat{a}(i)-1}\ \longrightarrow \mathbb{F}_{q^2}^{q^2}$ be a systematic encoder for a code $E_i$. The input sequence to the encoder $\phi_i$ are symbols from an array $\tilde{{\bf{r}}} = {\tilde{{\bf{r}}}_{i,j}}$ for $j = 0, 1, \dots, q^2 - \hat{a}(i) - 2$.

We want to construct an algorithm that takes as input an array d of size $q \times q^2$ with arbitrarily chosen symbols in positions ${(a,b): b = 0, 1, \dots, q-1; a = 0, 1, \dots, q^2 - \hat{a}(b) - 2}$ and zero in the remaining positions and that produces as output a code-array ${\bf{c}}$. Let $\hat{b}(j)$ be defined as the number of information symbols in the $j$th column of ${\bf{d}}$. The columns of ${\bf{d}}$ thus have the form ${\bf{d}}_j = ({\bf{d}}_{0,j}, {\bf{d}}_{1,j}, \dots,{\bf{d}}_{\hat{b}(j)-1}, 0, 0, \dots, 0)$.

We give a systematic encoder procedure in the following algorithm. During the procedure we also construct an array $\tilde{{\bf{r}}}$. The $i$th row of $\tilde{{\bf{r}}}$ is a codeword in $E_i$. Thus the first $q^2 - \hat{a}(i) - 1$ positions in the $i$th row of $\tilde{{\bf{r}}}$ determine the $i$th row of $\tilde{{\bf{r}}}$ completely.
\subsection{Algorithm 2}
The algorithm is shown in (\ref{algo2}).
\begin{figure*}
\hrulefill\\
\begin{equation}
\label{algo2}
Algorithm\ 2
\end{equation}
{\it{
Input: An $q \times q^2$ array ${\bf{d}}$. An empty $q \times q^2$ array $\tilde{{\bf{r}}}$.\par
Iterations: For $j = 0, 1, \dots, q^2 - 1$
\begin{enumerate}
\item Compute the known part of $\tilde{{\bf{r}}}_j$ for $i = 0, 1, \dots, q - 1 - \hat{b}(j)$ as 
\begin{displaymath}
\tilde{{\bf{r}}}_{i,j} = (\phi_i((\tilde{{\bf{r}}}_{i,0}, \tilde{{\bf{r}}}_{i,1}, \dots, \tilde{{\bf{r}}}_{i,q^2-\hat{a}(i)-2})))_j
\end{displaymath}
\item Solve the equation
\begin{displaymath}
A_j({\bf{d}}_{0,j}, {\bf{d}}_{1,j}, \dots, {\bf{d}}_{\hat{b}(j)-1, j}, y_{\hat{b}(j),j}, y_{\hat{b}(j)+1, j}, \dots, y_{q-1,j})^T = (\tilde{{\bf{r}}}_{0,j}, \tilde{{\bf{r}}}_{1,j}, \dots, \tilde{{\bf{r}}}_{q-1-\hat{b}(j),j}, u_{q-\hat{b}(j), j}, u_{q-\hat{b}(j)-1,j}, \dots, u_{q-1})^T 
\end{displaymath}
for $y_{\hat{b}(j),j}, y_{\hat{b}(j)+1, j}, \dots, y_{q-1,j}, u_{q-\hat{b}(j), j}, u_{q-\hat{b}(j)-1,j}, \dots, u_{q-1}$
\item Set
\begin{displaymath}
c_i = ({\bf{d}}_{0,j}, {\bf{d}}_{1,j}, \dots, {\bf{d}}_{\hat{b}(j)-1, j}, y_{\hat{b}(j),j}, y_{\hat{b}(j)+1, j}, \dots, y_{q-1,j})
\end{displaymath}
\begin{displaymath}
\tilde{{\bf{r}}}_i = (\tilde{{\bf{r}}}_{0,j}, \tilde{{\bf{r}}}_{1,j}, \dots, \tilde{{\bf{r}}}_{q-1-\hat{b}(j),j}, u_{q-\hat{b}(j), j}, u_{q-\hat{b}(j)-1,j}, \dots, u_{q-1})
\end{displaymath}
\end{enumerate}
}}
\vspace*{4pt}
\hrulefill
\end{figure*}

\newtheorem{theorem}{Theorem}
{\it{
\begin{theorem}
Algorithm 2 computes a code array $c$ of the Hermitian code $C = (ev_D(mP_\infty))^\bot$.
\end{theorem}
}}
\begin{proof}
The matrix $\tilde{{\bf{r}}}$ in the algorithm satisfies the conditions $\tilde{{\bf{r}}}_j$ = $A_j$ $c_j$ and the $i$-th row of $\tilde{{\bf{r}}}$ is a codeword in $E_i$. Thus $c$ is a code-array by Lemma 2.
\end{proof}

Algorithm 2 outlines the mathematical procedure to achieve systematic encoding of a Hermitian code. The real difficulty lies in an efficient implementation of the algorithm. We give such an implementation in Section IV but before proceeding we will need a simple lemma.

Let $A$ be any $n \times n$ matrix with inverse $A^{-1}$. We assume that the submatrix of $A$ indexed by elements $i, j, i = n - l, \dots, n-1\ and\ j = 0, 1, \dots, l - 1$ is nonsingular. This will always be true for the cases that we are interested in by Lemma 1. Let $I_l$ denote the $l \times l$ matrix and let $D(l)$ be a $n \times n$ matrix of the following form:\\
\begin{displaymath}
\mathbf{D(l)} = 
\left(\begin{array}{c|c}
I_l & 0 \\
\hline
P & 0
\end{array}\right)
\end{displaymath}
such that\\
\begin{displaymath}
\mathbf{A^{-1}D} = 
\left(\begin{array}{c|c}
0 & 0 \\
\hline
\tilde{P} & 0
\end{array}\right)
\end{displaymath}
for a $l \times l$ matrix $\tilde{P}$. We note that $D(l)^T$ is just a systematic encoding matrix for a code which has a parity check matrix the first $l$ rows of $A^{-1}$.
\begin{figure*}
\centering
\includegraphics[width=3.5in]{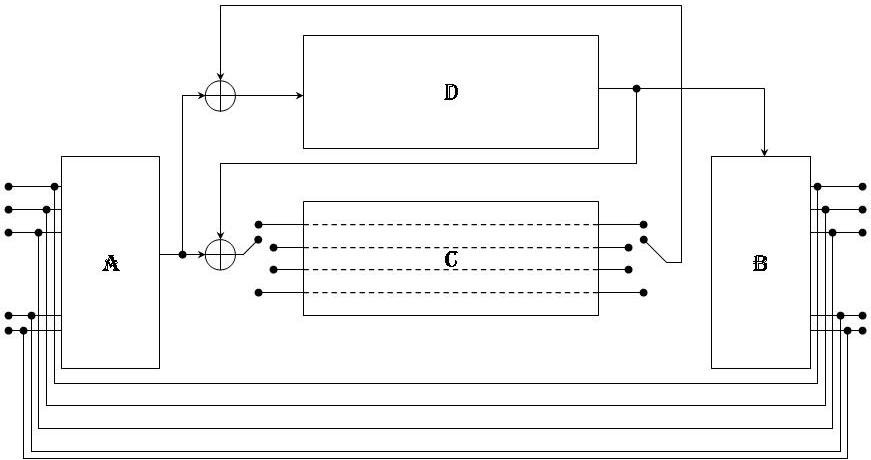}
\caption{Overall Outline of the Encoder Circuit. Switches $a$ \& $b$ are synchronized and rotate every clock cycle. The circuit is described in the text.}
\label{fig3}
\end{figure*}
\subsection{Lemma 5}
{\it{Let $x_0, x_1, \dots, x_{n-l-1}$ and $v_0, v_1, \dots, v_{l-1}$ be given. The solution for $y_{n-l}, y_{n-l+1}, \dots, y_{n-1}$ and $u_l, u_{l+1}, \dots, u_{n-1}$ to the linear system of equations
%
\setlength{\arraycolsep}{0.0em}
\begin{displaymath}
\begin{array}{ll}
A(x_0, x_1, \dots, x_{n-l-1}, y_{n-l}, y_{n-l+1}, \dots, y_{n-1})^T = &\\
\ \ \ \ (v_0, v_1, \dots, v_{l-1}, u_l, u_{l+1}, \dots, u_{n-1})^T &\\
\end{array}
\setlength{\arraycolsep}{5pt}
\end{displaymath}

can be found with the following algorithm.}}
\subsection{Algorithm 3}
Algorithm 3 is given as shown in (\ref{algo3}).
\begin{figure*}[hpb]
\hrulefill\\
\begin{equation}
\label{algo3}
Algorithm\ 3
\end{equation}
\begin{enumerate}
\item $b^T = A(x_0, x_1, \dots, x_{n-l-1}, 0, 0, \dots, 0)^T$
\item $\hat{b}^T = (v_0, v_1, \dots, v_{l-1}, 0, 0, \dots, 0)^T - (b_0, b_1, \dots, b_{l-1}, 0, 0, \dots, 0)^T$
\item $\tilde{b}^T = D\hat{b}^T$
\item $(x_0, x_1, \dots, x_{n-l-1}, y_{n-l}, y_{n-l+1}, \dots, y_{n-1})^T = (x_0, x_1, \dots, x_{n-l-1}, 0, 0, \dots, 0)^T + A^{-1}\tilde{b}$
\item $(v_0, v_1, \dots, v_{l-1}, u_l, u_{l+1}, \dots, u_{n-1})^T = \tilde{b}^T + b^T$
\end{enumerate}
\vspace*{4pt}
\hrulefill
\end{figure*}

\begin{proof}
$A^{-1}\tilde{b}^T = A^{-1}D\hat{b}^T$ which proves that $y_0, y_1, \dots, y_{n-l-1}$ equal zero.\\
Now it follows that\\
\begin{displaymath}
\begin{array}{ll}
& A (x_0, x_1, \dots, x_{n-l-1}, y_{n-l}, y_{n-l+1}, \dots, y_{n-1})^T \\
& = \tilde{b}^T + b^T\\
& = A(A^{-1}D\hat{b}^T) + b^T\\
& = D\hat{b}^T + b^T\\
& = D(v_0, v_1, \dots, v_{l-1}, 0, 0, \dots, 0)^T\\
&\ \ \ - D(b_0, b_1, \dots, b_{l-1}, 0, 0, \dots, 0)^T + b^T\\
& = (v_0, v_1, \dots, v_{l-1}, u_l, u_{l+1}, \dots, u_{n-1})^T\\
\end{array}
\end{displaymath}
\end{proof}
\section{Efficient Implementation of a Systematic Encoder}
\label{implement}
Inspecting Algorithm 2 and Lemma 4, we see that we need modules for multiplication of an array with matrix $A$, $A^{-1}$, systematic encoding of codes $E_j$, and a systematic encoding module for codes $D_l$ defined as
\begin{displaymath}
D_l = \{d \in \mathbb{F}_{q^2}^{q} : \sum_{j=0}^{q-1}A_{i,j}^{-1}d_j = 0, i = 0, 1, \dots, l-1\}
\end{displaymath}
Before describing the modules in detail we give a black box description and the overall description of the implementation.
\subsection{Module A: Multiplication with Matrix $A, A'$}
The module has as parallel input a vector $d$ of length $q$ and produces as serial output the numbers $(Ad^T)_i,\ i = 0, 1, \dots, q-1$ during the next $q$ clock cycles.
\subsection{Module B: Multiplication with Matrix $A^{-1}, A'^{-1}$}
Module B has as serial input a vector $d$ of length $q$. After $q$ clock cycles the parallel output is a vector $A^{-1}d^T$.
\subsection{Module C: Systematic Encoding of Codes $E_i$}
The module has a serial input of $q^2-\hat{a}(i)-1$ symbols and produce one symbol per clock cycle. The clocking frequency is $1/q$ of the overall clock rate.
\subsection{Module D: Systematic Encoding of Codes $D_l$}
Module D takes a serial input of length $l$ and produces as serial output a codeword of length $D_l$.
\subsection{Encoder}
Figure \ref{fig3} outlines the overall implementation. When the left hand input becomes valid, the output of Module A is added to the negative output of Module C, effectively implementing steps 1 and 2 of Algorithm 3. The sum is fed to Module D which implements step 3 of Algorithm 3. The output of Module D is combined with the output of Module A to implement step 4 of Algorithm 3. Simultaneously it is fed to module B of the implementation. After $q$ clock cycles the output of Module B is added to the input thus implementing step 5 of Algorithm 3.

Module C can be implemented as an obvious modification of a systematic encoding circuits for RS codes \cite{encode}. 

Module D implements systematic encoding of a code with parity check matrix given by first $l$ rows of matrix $A^{-1}$.
From the form of matrices $A^{-1}$, we see that code $D_l$ may be defined as 
\begin{displaymath}
\begin{array}{ll}
D_l = & \{d \in \mathbb{F}_{q^2}^{q} : \sum_{j=0}^{q-1}d_{q-1-j}(x_i)^j = d_0, x_0 = y_0, \\
&\ \ \ \ x_{s+1} = (y_0 + \gamma^s), s = 0, 1, \dots, l-2\}\\
\end{array}
\end{displaymath}
and we can use standard encoding techniques for shortened cyclic codes which are modified in the obvious way.
\section{Final Remarks}
\label{close}
A low complexity algorithm for systematic encoding and syndrome computation of Hermitian codes has been presented. The algorithm has a run time complexity of $O(n^{2/3})$ and is suitable for VLSI implementation. We give an outline for the encoder architecture, which uses as main blocks, $q$ varying-rate Reed Solomon encoders. The architecture achieves a much lower space complexity in terms of finite field multipliers and memory elements when compared to earlier reported works.
\section{Acknowledgement}
The authors would like to thank Science Foundation Ireland, Claude Shannon Institute and Deutsche Forschungsgemeinschaf for supporting parts of this research work.
%

\end{document}